# Graph-Theoretic Models of Resource Distribution for Cyber-Physical Systems of Disaster-Affected Regions

Kenneth Johnson, Samaneh Madanian
*Department of Computer Science Auckland
University of Technology Auckland
New Zealand*
{kenneth.johnson, sam.madanian}@aut.ac.nz

Roopak Sinha
*Department of IT & Software Engineering
Auckland University of Technology Auckland
New Zealand*
roopak.sinha@aut.ac.nz

**Abstract**

*We propose a tool-supported framework to reason about requirements constraining resource distributions and devise strategies for routing essential services in a disaster-affected region. At the core of our approach is the Route Advisor for Disaster-Affected Regions (RADAR) framework that operates on high-level algebraic representations of the region, modelled as a cyber-physical system (cps) where resource distribution is carried out over an infrastructure connecting physical geographical locations. The Satisfiable-Modulo Theories (SMT) and graph- theoretic algorithms used by the framework supports disaster management decision-making during response and preparedness phases. We demonstrate our approach on a case study in disaster management and describe scenarios to illustrate the usefulness of RADAR.*

## 1. INTRODUCTION

Disasters are hazards which may occur suddenly and in most cases without warning. They can cause loss of human life and damage to buildings and transportation infrastructure. Depending on disaster types and their severity, we can expect resources to be in short supply. Requirements of resource reallocation from other regions pose logistical issues [1]. Resolving complex requirements has been known as one of the dominate disaster management challenges [2]. To mitigate these risks, different organisations such as civil defence take great care in preparing for disaster scenarios through education and designing regional policies and plans to distribute resources of sustenance, medical supplies, survival equipment well in advance. It is believed that the root cause of the logistics problems in disaster response is not resource shortage but rather it is mostly about from failures to coordinate their distribution [3]. Therefore effective disaster response requires management and coordination of multiple organisations such as emergency services, and civil and technological agencies to minimise casualties and infrastructure damage.

Mathematical reasoning using graphs have been applied to solve various problems in serving disaster-affected regions [4]. Several works focus on resource distribution aspects, such as algorithms for identifying optimal placement of units for dis- patching and distributing resources [5], resource distribution in IoT based systems [6], and distributed constraint-solving [7]. Most of these solutions focus on specific resource types or depend on the availability of specialised technology, making it difficult to use them in more generalized settings.

One of the key *tasks* for the *Management Team (MT)* of a disaster is to devise and communicate *routes* for delivering critical resources or advising evacuation plans through a region affected by a disaster. A route is a finite sequence of connected locations through the region starting with a source and ending in a destination. There are associated costs based on distances between locations and routes must avoid damaged roadways or locations. Under rapidly changing conditions the Management Team must base their decisions on the most up to date knowledge of the disaster-affected region to have quality decisions regarding the available resources allocation [8]. On the other hand, still most activities and task associated with resource allocation are operated manually, which pose inefficiency in most cases [9].

To address these challenges, we formalise an algebraic specification of the *Route Advisor for Disaster Affected Regions (RADAR)* framework, which models the disaster-affected region as a cyber-physical system (cps). The management team uses RADAR to determine distributions of critical resources across a region based on requirements plans developed during the preparedness phase of disaster management. Requirements are automatically translated from a high-level language to first-order formulae and Satisfiable-Modulo-Theories (SMT) solving is used to compute a resource distribution satisfying the constraints. Appropriate amounts of resources are dis- tributed to locations across the region and this acts as an initial state of the cps. During a disaster, RADAR maintains a graph-theoretic model of the region and is programmable with graph algorithms to query the graph. For example, to test reachability between locations in the region or compute the shortest path to deliver essential services. To ensure queries are performed on models representative of an evolving and rapidly changing disaster scenario, RADAR integrates regional infrastructure availability updates and localised changes to resource requirement to transform the cps.

The main contributions of our paper are:

1) a graph theoretic model of a disaster-affected region
2) an adaptation of efficient SMT constraint solving for resource distribution across a region
3) a mathematical description of RADAR
4) a case study illustrating RADAR
5) a freely available open-source tool integrating RADAR with Microsoft's Z3 SMT Solver
6) preliminary experiments demonstrating RADAR.

The remainder of the paper is as follows. Section 2 gives a brief outline of the mathematical preliminaries needed to keep the paper self-contained. Section 3 introduces the RADAR framework and its *initialisation* and *continual analysis* phases. The Metropolis case study is described in Section 4. Section 5 gives an algebraic specification of the cps modelling the disaster-affected region and describes the system's key operations relating to disaster-management. Section 6 formalises local resource requirements and applies the approach to the Metropolis case study. Section 7 describes an approach to incremental constraints resolution which is adapted to resourcing constraints over a graph and illustrates with scenarios from the Metropolis case study. Section 8 outlines our implementation of RADAR and demonstrates its usage across a range of scenarios. Section 9 provides concluding remarks and describes future work.

## 2. PRELIMINARIES

Our work formalises resource distribution requirements as logical formulae, where variables store integer-values. The assignment of a value to a variable determines the amount of a particular resource in a particular geographical location. Let $F$ be the set of quantifier-free first order formulae over integer-sorted variables in $Z$ defined by the grammar

$$\phi ::= \phi_1 = \phi_2 \mid \phi_1 > \phi_2 \mid \phi_1 < \phi_2 \mid \neg \phi_1 \mid \phi_1 \wedge \phi_2$$

where $\phi_1$ and $\phi_2$ are formulae in $F$. When we need to specifically identify the variables appearing in a formula we write $\phi(z_1, ..., z_n)$, which means $\phi$ contains at least one occurrence of variables $z_i \in Z$, for $1 \leq i \leq n$. We also write $vars(\phi(z_1, ..., z_n)) = \{z_1, ..., z_n\}$ to denote the set of variables in $\phi$. Two formulae $\phi_1$ and $\phi_2$ are independent

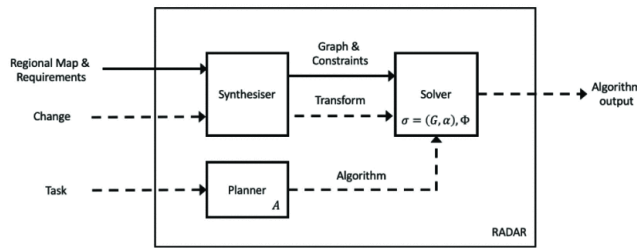

Fig. 1: Workflow of RADAR Framework

whenever they have no variables in common: $vars(\phi_1) \wedge vars(\phi_2) = \emptyset$.

### A. Formula Evaluation and Satisfiability

Let $\alpha: Z \to \mathbb{Z}$ be an assignment such that the equation $\alpha(z) = q$ assigns the integer value $q \in \mathbb{Z}$ to the variable $z \in Z$. Let $[Z \to \mathbb{Z}]$ be the set of all such maps. An assignment $\alpha \in [Z \to \mathbb{Z}]$ is extended to formula evaluation $[\![-]\!] : F \times [Z \to \mathbb{Z}] \to \mathbb{Z}$ such that $[\![\phi(z_1, ..., z_n)]\!] (\alpha)$ works out the value of $\phi$ from the variable assignment $\alpha(z_1), ..., \alpha(z_n)$. We say the formula $\phi$ is satisfiable if, and only if, there exists an assignment $\alpha: Z \to \mathbb{Z}$ such that $[\![\phi]\!] (\alpha) = true$. In this case, we say that $\alpha$ satisfies $\phi$ and we write $\alpha \vDash \phi$.

A useful operation for our work is the $update: [Z \to \mathbb{Z}] \times Z \times A \to [Z \to \mathbb{Z}]$ operation, which modifies an assignment α at a single variable $z \in Z$ and is defined by $update(\alpha, z', q)(z) = q$ if $z = z'$ and $\alpha(z)$ otherwise, for the value $q \in \mathbb{Z}$ and $z' \in Z$.

## 3. THE RADAR FRAMEWORK

The workflow for the *Route Advisor for Disaster-Affected Regions (RADAR)* framework is presented in Figure 1. The framework is used by the Management Team (MT) to work out a resource distribution compliant with requirements devised during the disaster preparedness phase which could be challenging as it is associated with a high level of uncertainty [10]. RADAR automatically applies graph algorithms for advising resource distribution and evacuation routes across an evolving disaster-affected region scenario.

In this section, we provide an overview of the RADAR framework and its usage. RADAR has two phases. The *initialisation phase* constructs an initial state $\sigma = (G, \alpha)$ of the cyber-physical system modelling the disaster-affected region, where $G$ is a graph model of the region and $\alpha$ a resource distribution assignment. During the *continual analysis phase*, tasks are input to RADAR to be analysed against the current state of the cps. Updates observed from the disaster-affected region are translated into transformations on the system's state and reflected in future tasks.

### A. Initialisation Phase

Figure 1 represents inputs accepted to RADAR during the *initialisation phase* as solid lines. In this phase, the *synthesiser* takes as input a regional map including distances between locations within the disaster-affected region and synthesises a weighted, directed graph $G$. This step may be automated and graphs synthesised from geodata available in online repositories. Resources distribution policies are formulated by agencies such as the National Emergency Management Agency (NEMA) in New Zealand according to preparedness guidelines. The requirements map $\Phi$ for $G$ is translated from high-level resource allocation requirement devised during disaster preparedness.

Both the synthesised graph $G$ and requirements map $\Phi$ are provided to the Solver component. The Solver utilises the Satisfiability modulo theories (SMT) tool *smt* to work out an assignment $\alpha : Z \rightarrow \mathbb{N}$ that determines quantities of resources such that $\alpha \vDash \Phi$. RADAR sets the initial state as $\sigma = (G, \alpha)$.

### B. Continual Analysis Phase

When RADAR's initialisation is complete, the *continual analysis phase* begins. Inputs to RADAR during this phase are represented by dashed lines in Figure 1 and are provided continually as the disaster scenario affects and changes the region. During this phase, the disaster management team inputs *tasks* expressed in a high-level natural language and are translated by the *Planner* to a cps operation defined in Section V-C:

1) $reach(\sigma, s, t)$: determines reachability of $t$ from $s$ by checking if a route between the locations exists

2) $route(\sigma, s, t)$: computes the shortest route between $s$ and $t$ based on distance weights in the graph $G$

3) $deliver(\sigma, s, t, q)$: delivers q resources from $s$ to $t$ using the shortest route between the locations. The operation is performed only if such a route exists.

RADAR implements each operation via an algorithm contained in the Planner's library A of standard graph algorithms. The Solver executes the algorithm on the graph G and out- puts the task's algorithm *output* for further analysis by the disaster management team. This component of the proposed framework addresses making informed decisions on resource allocation during dynamic response [11].

### C. Updating the State

To ensure tasks are completed on a graph model aligned with the region's current status, the synthesiser receives notifications of situational changes within the region. This component of the framework is addressing the requirement stated by [12] to incorporate important features and changes in resource management software. In this paper, we consider two kinds of changes: (i) availability of locations or streets and (ii) modification of resource distribution requirements.

## 4. METROPOLIS CASE STUDY

Figure 2 presents a regional map of the fictitious city of Metropolis with 7 locations representing important landmarks, labelled alphabetically from a to g. Each two-way street on the map has a name label and the distance between locations given in bold underline. Some streets and locations are labelled with special icons. The ⚓ icon at location a represents a shipping port which receives and properly stores fresh supplies during disaster scenarios. The

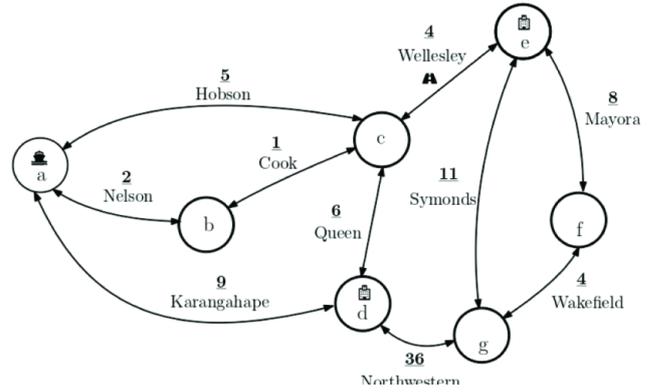

Fig. 2: Regional map of the fictional city of Metropolis

🏥 icon at locations d and e represent large medical centres. Wellesley street is labelled by the 🌉 icon representing a bridge over a salt-water bay separating locations c and e. Alternatively, a route exists between c and e by travelling around the bay, via the Northwestern motorway.

To facilitate an appropriate disaster response, resources must be distributed in a way satisfying guidelines and plans set out during the preparedness phase. This could substantially help the management team to get decisions under uncertain and complex situations, especially when they are dealing with resource constraints [13]. To simplify our discussion, we suppose only three sorts of resources need to be distributed during a disaster scenario: sustenance **S**, medicine **M** and survival equipment **E**. Each sort of resource is counted and distributed across locations by number of *units*. We distribute these resources according to the following requirements:

**R1** shipping location a has 100 units of medicine, sustenance and equipment

**R2** locations connected via the bridge have the same units of sustenance

**R3** hospitals receive at least 20 units of medicine, 15 of sustenance and 23 units of equipment

**R4** location g receives between 7 and 10 units of medicine

**R5** location f receives 5 more units of medicine and sustenance than location g

**R6** locations within distance 8 of a receives at least 40 units of sustenance

**R7** Only hospital or supply locations can receive equipment resources.

A *resource distribution* associates a specific amount of medicine, sustenance and equipment to each location and satisfies requirements R1 to R7. A resource distribution for locations in Metropolis is depicted in Table I.

TABLE I: Resource Distribution across Metropolis, satisfying requirements R1 to R7.

|   | a   | b  | c  | d  | e  | f  | g |
|---|-----|----|----|----|----|----|---|
| **S** | 100 | 40 | 40 | 16 | 40 | 5  | 0 |
| **M** | 100 | 0  | 0  | 21 | 21 | 12 | 7 |
| **E** | 100 | 0  | 0  | 24 | 24 | 0  | 0 |

# 5. MODELLING DISASTER-AFFECTED REGIONS AS CYBER-PHYSICAL SYSTEMS

In this section, we model a *cyber-physical system (cps)* as an algebra comprising a set $\Sigma = \{f_1, \ldots, f_n\}$ of operations modelled by functions over one or more carrier sets of data. We typically assume our data includes the Booleans $\mathbb{B}$ in order to define *tests*; operations in $\Sigma$ of the form $f : A \to \mathbb{B}$. Our work models a cyber-physical system as an algebra $(\Sigma, State)$ such that $State$ is the set of all cps states and $\Sigma$ is a finite set of operations and tests on states in State. We define the operations below. For operations of the form $f : State \to State$, the equation $f(\sigma) = \sigma'$ modifies the state and $\sigma'$ is the state that results from the application of operation $f \in \Sigma$ on $\sigma$. In this section, we define the cps state as the product $State = G \times [Z \to \mathbb{Z}_{\geq 0}]$. such that the pair $\sigma = (G, \alpha) \in State$ comprises a graph $G$ and resource distribution assignment $\alpha$.

## A. Graph Representation of a Region

Let graph $G = (V, E)$ be obtained from a regional map such that $V$ is the finite set of vertices representing locations on the map and $E = V \times V$ is the finite set of weighted edges between vertices. For the two-way street of distance $w$ connecting locations $v$ and $v'$ in $V$, define a bidirectional edge in $E$ such that $v \neq v'$ and $(v, v') \in E \iff (v', v) \in E$, both with weight $w$. Each graph element $v \in V$ and $e \in E$ has a Boolean-valued *availability* status which models its availability.

*Metropolis Graph:* Figure 2 presents the regional map of Metropolis. We define the graph $G = (V, E)$ in a straightforward way, setting vertices $V = \{a, \ldots, g\}$ to correspond to locations in the map. Next, define two weighted directed edges in $E$ for each two-way street for a total of 18 edges. For example $(a, b)$ and $(b, a)$ in $E$ are the pair of edges representing Nelson street, such that weight $w = 2$ for edges $(a, b)$ and $(b, a)$. Initially, all Metropolis graph elements are available.

## B. Resource Distribution Models

Each vertex $v$ is associated with a finite subset $Z_v = \{vz_1, \ldots, vz_m\}$ of variables in $Z$ such that each $vz_i$ is assigned a value representing a specific type of resource at location $v$, for $1 \leq i \leq m$. We call $Z_v$ the *model* of $v$ and the elements of $Z_v$ resource variables. Clearly, the set $Z$ of variables is partitioned over vertices from $V$ such that $Z_v \cap Z_{v'} = \emptyset$ for $v \neq v'$.

An induced subgraph $G[S]$ of $G$ is a graph containing vertices of $G$ such that $S \subseteq V$. The edges of $G[S]$ is any edge of $G$ whose endpoints are vertices in $S$. Let $SG$ be all induced subgraphs of the graph $G$. We extend the notion of a model to any induced subgraph $G[S]$ of $G$ by setting $Z_S = \bigcup_{v \in S} Z_v$.

*Metropolis Models:* We consider models for location vertices in Metropolis. Each $v \in V$ is modelled by the set $Z_v = \{\mathbf{vS}, \mathbf{vM}, \mathbf{vE}\}$ storing integer values that measure quantities of sustenance **vS**, medicine **vM** and equipment **vE.** Now, the model for the induced subgraph comprising vertices $S = \{c, e\}$ is the union of both models for c and e. In symbols, $Z_S = Z_c \cup Z_e = \{\mathbf{cS}, \mathbf{cM}, \mathbf{cE}\} \cup \{\mathbf{eS}, \mathbf{eM}, \mathbf{eE}\}$.

## C. Tests and Operations

A disaster-affected region is a dynamically evolving scenario. Adverse events disrupt and affect locations in the region, depleting resources and causing damage to infrastructure such as roadways. Therefore dynamic information collection and resource tracking is an essential components of resource management system. To model such updates, we define simple operations and tests that change the states of the cps. Throughout, let $\sigma = (G, \alpha)$ be a state in $State$.

*Route Operation:* The operation $route : State \times V \times V \to Path$ is defined such that the equation $route(\sigma, s, t) = p$ means that $p = v_1 \to v_2 \to \cdots \to v_n$ is a path or route through $G$ comprising vertices $v_i \in V$ for $1 \leq i \leq n$ such that $v_i = s$, $v_n = t$ and there exists an edge between each vertex in the path. Each graph element of $p$ must be available in the state $\sigma$. If no such route exists then $route(\sigma, s, t) = \mathbf{u}$, where $\mathbf{u} \notin Path$ is a special element to denote the operation was unsuccessful.

*Test Graph Reachability:* The operation $reach : State \times V \times V \to \mathbb{B}$ is a test such that $reach(\sigma, s, t) = false$ if $route(\sigma, s, t) = \mathbf{u}$ and $true$ otherwise, for any two location vertices $s, t \in V$.

*Set Graph Element Operation:* Let $g \in V \cup E$ be a graph element; either a vertex in $V$ or an edge in $E$. Let $set : State \times G \times \mathbb{B} \to State$ be an operation such that $set(\sigma, g, b) = \sigma'$ resulting in the state $\sigma'$ with availability value of $g$ set to $b$.

*Test Graph Element Availability:* We may determine the availability of a graph element in $G$ by defining the test $available : State \times G \to \mathbb{B}$ such that $available(\sigma, g) = b$ means the graph element $g$ of $G$ has availability $b$ in state $\sigma$.

*Audit Operation:* When resources arrive on location, staff carry out an audit. Depending on the sort of resource, auditing involves recording quantity and verifying its condition before storing for use. The outcome of an audit results in a transcript which provides salient details of the resources. Transcripts obtained from an audit at each location are stored, giving a snapshot of the resources available there. Let $Transcript$ denote the set of all audit transcript results. We formalise the auditing process at location vertex $v \in V$ via the operation $audit : [Z \to Transcript] \times Z_v \times Transcript \to [Z \to Transcript]$ such that the equation

$$audit(\alpha, z, t) \coloneqq update(\alpha, z, t) \qquad (1)$$

means an audit of resources at $v$ ensures the transcript $t \in Transcript$ is stored in the resource variable $z \in Z_v$ by updating the assignment map $\alpha$ of state $\sigma$, accordingly. All other variables are left unmodified.

A key aspect of the auditing process is determining the number of usable resource units at a location. Since our work only concerns the quantity of resources at a location, we can discard any other information from the transcript.

Mathematically, we simply set $Transcript = \mathbb{N}$ such that Equation (1) is then interpreted as the operation $audit : [Z \to \mathbb{N}] \times Z_v \times \mathbb{N} \to \mathbb{N}$.

*Delivery Operation:* Lastly, we consider an operation to deliver an amount of $q$ from a source vertex $s \in V$ to a target vertex $t \in V$ in the state $\sigma = (G, \alpha)$. The deliver is only possible when $t$ is reachable from $s$ in $G$. Suppose the resource to deliver is modelled by resource variables $sz \in Z_s$ at location $s$ and $tz \in Z_t$ at location $t$. We define $deliver: State \times Z_s \times Z_t \times \mathbb{N} \to State$ by the equation $(G, \alpha') = deliver(\sigma, sz, st, q)$ such that if $reach(\sigma, s, t) = true$ then the assignment mapping $\alpha'$ satisfies equations

$$\alpha'(sz) = \alpha(sz) - q$$
$$\alpha'(tz) = \alpha(tz) + q.$$

Otherwise, the state $\sigma$ is unmodified and $\alpha' = \alpha$.

In summary, the cps modelling a disaster-affected region is an algebra comprising states of the form $\sigma = (G, \alpha)$ and tests and operations in the signature $\Sigma = \{route, reach, set, available, audit, deliver\}$.

# 6. RESOURCE DISTRIBUTION CONSTRAINTS

As suggested by [14], recognizing the constraint of resources in disaster response is critical and we have incorporated this requirement into our proposed framework. In this section, we devise constraints that specify ranges of values that the assignment mapping may associate to the variables of location models. These requirements stem from disaster preparedness documentation and act as an *initial state* for the cps modelling a disaster-affected region.

## A. Requirement Map
We define the *requirement map* of the graph $G$ which assigns a logical formulae to induced subgraphs $G[S]$. The formulae constrain the models of locations in $S$ according to resource distribution requirements. In particular, we specify induced subgraphs $G[S_1], \ldots, G[S_p]$ that partition the vertex set $V$ of $G$ such that

$$S_i \wedge S_j = \emptyset \qquad (2)$$

for $i \neq j$ and $1 \leq i, j \leq p$.

We define $\Phi : SG \to F$ such that the formula $\Phi(G[S_i])$ constrains variables in the model $Z_{S_i}$. This formula has the property that it may only contain variables from the model $Z_{S_i}$. In symbols: $vars(\Phi(G[S_i])) \subseteq Z_{S_i}$, for $1 \leq i \leq p$. We suppose that each formula $\Phi(G[S_i])$ is in conjunctive normal form: e.g. $\Phi(G[S_i]) \equiv \wedge_j \phi_j$ of logical formulae $\phi_j$. We make $\Phi$ a total mapping by assigning the empty formula $\epsilon \in F$ to all other subgraphs in the domain of $\Phi$. The empty formula is satisfiable by all assignment mappings.

## B. Resource Distribution Constraints of Metropolis
We synthesise formulae for Requirements R1 to R7 and construct the mapping $\Phi$ for the Metropolis case study. For Requirement R1, we form the subgraph $G_1 = a$ comprising all supply location vertices . We set

$$\Phi(G_1) \coloneqq (\mathbf{aS} = 100) \wedge (\mathbf{aM} = 100) \wedge (\mathbf{aE} = 100). \qquad (3)$$

For Requirement R2, we form subgraph $G_2 = \{c, e\}$ with location vertices connected via a street edge labelled with ⛰. We set

$$\Phi(G_2) \coloneqq (\mathbf{cS} = \mathbf{eS}). \qquad (4)$$

Requirement R3 specifies constraints for resourcing hospitals: location vertices D and E. Now, for D we form subgraph $G_3$ such that

$$\Phi(G_3) \coloneqq (\mathbf{gM} \geq 20) \wedge (\mathbf{gS} \geq 15) \wedge (\mathbf{gE} \geq 23). \qquad (5)$$

However, we note that $e$ is already included in subgraph $G_2$. To maintain subgraph partition Property (2), we define the formula $\phi_e \equiv (\mathbf{eM} \geq 20) \wedge (\mathbf{eS} \geq 15) \wedge (\mathbf{eE} \geq 23)$ and update $\Phi$ such that

$$\Phi'(G_2) \coloneqq \Phi(G_2) \wedge \phi_e. \qquad (6)$$

Requirement R4 specifies resourcing constraints for location vertex $g$ and we define the subgraph comprising g

$$\Phi(G_4) \coloneqq (\mathbf{gM} \geq 7) \wedge (\mathbf{gM} \leq 10). \qquad (7)$$

Requirement R5 specifies resourcing constraints for location vertex f in terms of g. Maintaining partition property (2) we define the formula $\phi_f \equiv (\mathbf{fM} = \mathbf{gM} + 5) \wedge (\mathbf{fS} = \mathbf{gS} + 5)$ and update $\Phi$ such that

$$\Phi'(G_4) \coloneqq \Phi(G_4) \wedge \phi_f. \qquad (8)$$

Requirement R6 identifies location vertices b and c which are within a distance of 8 from supply a according to Metropolis' street edge weights. We define $\phi_b \equiv (\mathbf{bS} \geq 40)$ and $\phi_c \equiv (\mathbf{cS} \geq 40)$ to construct subgraph $G_6$ with location vertex $b$, setting

$$\Phi(G_6) \coloneqq \phi_b. \qquad (9)$$

As c is already in subgraph $G_2$, we update

$$\Phi'(R_2) \coloneqq \Phi(R_2) \wedge \phi_c. \qquad (10)$$

Requirement R7 states that only hospitals may have equipment on-site. In fact, this is a constraint on every location not identified as a hospital  or supply  : namely b, c, g, f. Since each of these location vertices are contained in a previously defined subgraph we will need to update $\Phi$ accordingly:

$$\Phi'(G_6) = \Phi(G_6) \wedge (\mathbf{bE} = 0)$$
$$\Phi'(G_2) = \Phi(G_2) \wedge (\mathbf{cE} = 0)$$
$$\Phi'(G_4) = \Phi(G_4) \wedge (\mathbf{gE} = 0) \wedge (\mathbf{fE} = 0)$$

Finally, since there cannot be a negative amount of resources in any location, we form a default constraint such that $\phi_v^d \coloneqq \mathbf{vS} \geq 0 \wedge \mathbf{vM} \geq 0 \wedge \mathbf{vE} \geq 0$ for any vertex $v \in \{a, \ldots, f\}$. Then for each vertex $v$ in each subgraph $G_1, \ldots, G_6$ we update the requirements mapping such that $\Phi'(G_i) \coloneqq \Phi(G_i) \wedge \phi_v^d$. To ensure $\Phi$ is total, we set $\Phi(G[S]) \coloneqq \epsilon$ for any other induced subgraph $G[S]$ of $G$.

# 7. INCREMENTAL CONSTRAINT RESOLUTION

By translating resource distribution requirements to logical constraints, we wish to utilise automatic approaches via SMT solvers to obtain a satisfiable assignment to compute resource quantities per location in a region.

An SMT (Satisfiability Modulo Theories) solver takes as input a logical formula $\phi$ in $F$ containing variables in $Z$ and works out an assignment $\alpha : Z \to A$ that assigns values from a set $A$ to variables of $Z$ such that the formula is evaluated to $true$. We model this mathematically by $smt : F \to [Z \to A] \cup \{u\}$ such that $smt(\phi) = \alpha$ if, and only if, $[\![\phi]\!](\alpha) = true$. In symbols we write $\alpha \vDash \phi$. If no such $\alpha$ exists, then $\phi$ is unsatisfiable and we write $smt(\phi) = u$, for special element $u \notin F$. However, we can expect even a modestly sized region to be modelled by a graph with several thousand vertices. If the number of vertices is $n$ and each vertex has a model with $q$ variables, then the state space of variables is $n \cdot q$. While SMT solving technologies continue to improve and widen the scope to larger and more complex constraints, we can quickly reach intractable problems. Furthermore, we can expect localised requirements of the region to evolve in complex, unexpected ways during a disaster. Intuitively, we should only resolve this locality to determine its needs, while resourcing for the rest of the region remains unchanged.

## A. Compositional Resolution of Constraints

To address these technical challenges, we adapt the compositional SMT constraint resolution approach from [15] to resolve constraints over a graph model. This approach exploits the fact that for two independent formulae $\phi_1$ and $\phi_2$ if $\alpha_1 \vDash \phi_1$ and $\alpha_2 \vDash \phi_2$ then

$$(\alpha_1 \oplus \alpha_2) \vDash \phi_1 \wedge \phi_2. \quad (11)$$

where the composition operation $\oplus : [Z \to \mathbb{Z}] \times [Z \to \mathbb{Z}] \to [Z \to \mathbb{Z}]$ is defined as $(\alpha_1 \oplus \alpha_2)(z) = \alpha_1(z)$ if $z$ is in the domain of $\alpha_1$, and $\alpha_2(z)$ otherwise. This property extends to $n$ independent by an inductive argument: if $\alpha_i \vDash \phi_i$ for $1 \leq i \leq n$ then $\oplus_{i=1}^{n} \alpha_i \vDash \wedge_{i=1}^{n} \phi_i$.

We outline a compositional approach to resolving graph constraints in the following steps:

1) Partition $\phi$ into pairwise independent formula $\phi_1, \ldots, \phi_p$ such that $\phi = \wedge_{1 \leq j \leq p} \phi_j$
2) Compute $\alpha_j = smt(\phi_j)$, for $1 \leq j \leq p$
3) Construct $\alpha = \oplus_{j=1}^{p} \alpha_j$.

These steps ensure $\alpha \vDash \phi$.

We apply compostional SMT resolution to compute a satisfiable assignment map $\alpha$ for constraints over a graph $G$ as specified by requirement mapping $\Phi : SG \to F(\Gamma, Z)$. First, we set $\phi \equiv \wedge_{i=1}^{p} \phi_i$ to be the formulae to resolve such that $\phi_1 \equiv \Phi(G[S_1]), \ldots, \phi_p \equiv \Phi(G[S_p])$. Clearly, the $\phi_i$'s are independent formulae by Property (2). Hence, we obtain a satisfiable $\alpha_i \vDash \phi_i$ for each one in Step 2. Step 3 constructs the satisfiable assignment mapping such that $\alpha \vDash \phi$ we required.

By applying compositional resolution to the Metropolis constraints synthesised in Section 6-B, we obtain a satisfiable assignment of Requirements R1 to R7 as shown in Table I.

## B. Incremental Re-resolution of Constraints

During a disaster scenario, constraints in $\Phi$ affecting one or more locations may be modified by the management team. Clearly, local constraint modifications do not affect an entire region and so instead of re-resolving a new resource distribution $\alpha'$ from scratch, we selectively resolve only those constraints affected by a change. To this end, we describe steps to compute a new satisfiable assignment mapping $\alpha'$ after a change in the requirement map.

To this end, we define the operation $transform : [SG \to F] \times F \to [SG \to F]$ on requirement maps such that $transform(\Phi, \phi(z_1, \ldots, z_m))$ identifies vertices affected by a supplied formula $\phi(z_1, \ldots, z_m)$ and updates $\Phi$ accordingly. Special care is taken to ensure the transformation maintains the independent formulae property (11). We construct this updated requirement map $\Phi'$ in the following incremental re-resolution steps:

1) Identify vertices $v_1, \ldots, v_m$ affected from addition of $\phi$ by noting $\phi(z_1, \ldots, z_n)$ has variable occurrences $z_1, \ldots, z_n$ from resource models $M_{v_1}, \ldots, M_{v_m}$.

2) Set $S \coloneqq \cup_{i=1}^{m} G_i$ from the corresponding affected subgraphs $G_1, \ldots, G_m$ such that $v_i \in G_i$ for $1 \leq i \leq m$

3) There are two ways to update the constraints for $S$:

   - Set $\Phi'[S] \coloneqq \wedge_{i=1}^{m} \phi_i \wedge \phi$ such that $\phi_i = \Phi[G_i]$, which includes all existing formulae constraining vertices from affected subgraphs in the previous step and adds the new formula $\phi$
   - Set $\Phi'[S] \coloneqq \phi$, which replaces all existing formulae constraining vertices from affected subgraphs in the previous step with the new formula $\phi$.

We reuse value assignments of all unaffected vertices to define a satisfiable assignment map $\alpha' \vDash \Phi'$ by computing $\alpha_S = smt(\Phi'(G[S]))$ and setting $\alpha' = \alpha_S \oplus \alpha$.

## C. Scenario: Updating Metropolis Supply Constraints

As a disaster scenario evolves, Metropolis' requirements defined in Section 6-B are modified to admit a broader range of acceptable resource distributions. Metropolis supply location a ensures that there are 100 audited units of sustenance, medicine and equipment. As resources are distributed, it is important that a stock of 50 units of each resource remains on-site. Translating the high-level requirements to a logical formula, we define

$$\phi_a \equiv (\mathbf{aM} \geq 50) \wedge (\mathbf{aS} \geq 50) \wedge (\mathbf{aE} \geq 50).$$

To apply the operation $transform(\Phi, \phi_a)$ we follow the incremental re-resolution steps:

1) The only location affected by the change is a, since the variables of $\phi_a$ are contained in $M_a$
2) Set $S := G_1$ since the corresponding subgraph containing $a$ is $G_1$
3) Set $\Phi'(S) := \phi_a$, Replacing existing constraints for a with the new ones

Compute $\alpha S = smt(\phi_a)$ and form $\alpha' = \alpha S \oplus \alpha$ such that $\alpha' \vDash \Phi'$. Assignment $\alpha'$ sets

$$\alpha'(\mathbf{aS}) = 50 \, \alpha'(\mathbf{aM}) = 50 \text{ and } \alpha'(\mathbf{aE}) = 50$$

and no other variable changes its assigned value.

### D. Scenario: Increasing Localised Aid Capacity

In this scenario, we update constraints for locations b and c such that both locations have at least 50 units of sustenance and both have identical amounts of at least 10 units of medicine. Translating this high-level requirement to a logical formula, we obtain

$$\phi_{bc} \equiv (\mathbf{bS} \geq 50) \wedge (\mathbf{cS} \geq 50) \wedge$$
$$(\mathbf{bM} \geq 10) \wedge (\mathbf{cM} \geq 10) \wedge$$
$$(\mathbf{bM} = \mathbf{cM}).$$

We apply the operation *transform*($\Phi$, $\phi_{bc}$) by following the incremental re-resolution steps:

1) The affected vertices are b and c since the occurrences in $\phi_{bc}$ have variable from resource models $M_b$ and $M_c$
2) Set $S := G_2 \cup G_6$ since the corresponding subgraphs are $G_2 = \{c, e\}$ and $G_6 = \{b\}$
3) Set $\Phi'[S] := \phi_2 \wedge \phi_6 \wedge \phi_{bc}$ to include all existing formulae constraining b, c and e such that $\phi_2 = \Phi[G2]$ and $\phi_6 = \Phi[G_6]$ and add the new formula $\phi_{bc}$.

Compute $\alpha_S = smt(\phi_S)$ and form $\alpha' = \alpha_S \oplus \alpha$ such that $\alpha' \vDash \Phi'$.

The assignment $\alpha_S$ sets values for the variables corresponding to the amounts of sustenance, medicine and equipment at locations b, c and e as follows:

|   | b | c | e |
|---|---|---|---|
| **S** | 50 | 50 | 16 |
| **M** | 10 | 10 | 21 |
| **E** | 0 | 0 | 24 |

All other variables maintain their previous value assignment.

## 8. IMPLEMENTATION AND EXPERIMENTS

We implemented RADAR as part of an open-source Java library that is freely available from Github[1]. The `Radar` class maintains instances of all key components of the framework: `Syntheiser`, `Planner` and `Solver` and exposes the following public methods:

- `initialise()` is invoked to initialise the Radar object and corresponds to the framework's initialisation phase. The `Syntheiser` translates resourcing constraints over the input graph model of the region and invokes `Solver` to compute a satisfiable assignment mapping amounts of resources to each location using SMT solver Z3 [16]
- `compute(Task)` implements continual analysis of tasks by invoking the `Planner` to compute a result for each of the following concrete subclass extension of `Task`: `Reachable`, `Route` and `Deliver`. An Audit is performed after completing a `Deliver` task and changes quantities of resources in the graph
- `update(Update)` invokes `Syntheiser` to update the graph or resource constraints, taking as a parameter any extension of the abstract Update class. We implemented two extensions: `Availability`, updates a graph element according to a new availability status and `LocaleResource` which invokes `Solver` to resolve an update in resource constraints.

An instance of `Radar` is created by constructor public `Radar(RegionMap,Requirements)` such that abstract classes RegionMap and Requirements describe the region and specify resourcing requirements. Both objects are input to the `Synthesiser` which outputs a graph and logical formulae for constraints. The graph is a directed weighted graph implemented by the `JGraphT` library [17] with `Location` objects as vertices and `Street` objects for edges. `Location` objects maintain instance variables that store resource amounts. Both classes may be extended to store detailed information such as population density or road capacity. The `Planner` component invokes JGraphT's implementation of the Bellman-Ford Shortest Path algorithm to determine a suitable route each task. Algorithms are run on a subgraph of the region containing only those vertices and edges whose status indicates they are available.

We instantiated Metropolis' regional map in Figure 2 as a graph comprising 7 vertices and 20 edges. Using the requirements map defined in 7, the initialisation process took about 79 milliseconds to complete and computed the satisfiable resource distribution assignment presented in Table I. We illustrate RADAR's use by invoking `compute` on the following tasks objects:

- `Reachable(a,f)` to RADAR which determines if there exists a route between locations a and f, which returns `true`.
- `Route(a,f)` task to determine the path between a to f compute the shortest path (a → b → c → e → f) with a weight of 15.0

---
[1] https://github.com/emsoftaut/2020-RADAR

- `Deliver(a,f,S,M,E)` where setting $S = 12, M = 12$ and $E = 0$ delivers quantities of sustenance, medicine and equipment from the supply located at a to location f As a result, the quantities at the supply are decreased. The task is completed using the shortest path as in the previous routing task and the returned `Audit` object confirms an updated amount of 22 sustenance, 17 medicine and 0 equipment at location f.

Now, we suppose that it is observed the Wellesley Bridge in Metropolis becomes unavailable due to damage causing instabilities in the bridge. We update the graph model by constructing an appropriate `Availability` object and invoke the `update` method, which sets the status of the corresponding street edges c → e and e → c to false. Computing the route between locations a and f again yields the shortest path (a → d → g → f) with the weight 49.0, which avoids unavailable graph elements and instead includes Northwestern between d and g.

## 9. CONCLUSION AND FUTURE WORK

This paper proposes the use of a graph-theoretic model of cyber-physical systems and SMT solving to address the challenge of distributing critical resources in a disaster-affected area. Resource distribution is one of the features of the proposed RADAR framework, which flexibly supports solving other problems like finding shortest-paths for evacuations and identifying network capacity and flow bottlenecks for mass transits. The framework is able to support real-time decision-making by adapting its solutions to updated knowledge about the region. Optimisations such as incremental re-resolution allow us to localise the impact of changes to only a subset of the decisions that have already been made.

Our approach treats disaster-affected regions as a cyberphysical system that comprises a network of physical location components over which a satisfiable resource distributed is computed. We modelled system tests that sense availability of physical components of the graph to deliver resources based on this availability. We can extend the RADAR framework by modelling the probabilistic in nature of location and path availability. Uncertainties may be based on unstructured data obtained from social media platforms such as in [18] and also from measuring the reliability of the network's underlying physical infrastructure, such as roadways. Uncertainties also arise from human operators who carry out key operations to deliver and audit resources. Systems that explicitly modelling human-in-the-loop behaviour are classed as socio-cyberphysical systems [19]. Human participants carrying out tasks in the system can be formally modelled using the Opportunity-Willingness-Capability ontology [20].

At this stage, the research is in the preliminary phase and the developed software is a proof of concept. The future directions of this work would be involving with the potential end-users/stakeholders to further investigate and identify the most important criteria for the resource distribution from their perspective. This would help us to add the related features to the software and solving resource distribution problems in unstable situations more efficiently. Also, identifying these criteria based on the stakeholders' perspective gives us a solid foundation in defining proper Key Performance Indicators (KPIs) to scientifically compare our proposed model with the available methods and approaches.